Title: Semi-Circular Canals Anomalies//Idiopathic Scoliosis

Authors: D. L. Rousie (1), J.P. Deroubaix (2), O. Joly (1), P. Salvetti (2), J. Vasseur (2), A. Berthoz (1), ( (1) Laboratoire de la perception et de l action - College de France - Paris, (2) IFR 49 INSERM/CNRS - Orsay)

Comments :

Name and adress for correspondence: Docteur Rousie, 3 rue Saint Louis, 59113 Seclin France. Fax: 0033 320 32 35 44, Tel. 0033 320 90 12 29, mrousie@nordnet.fr

Supports:

Key points:

- Membranous semi circular canals (SCC) modelling based on MRI revealed significant anomalies in IS patients compared to normal subjects
- Frequent aplasias located in the lateral canal were found in IS.
- We also discovered a, never described, abnormal communication between lateral and posterior canal.
- Lateral SCC is involved in trunk rotation and lateral deviation: these movements are frequently abnormal in IS.

Structured Abstract:

Study design

Are there qualitative and/or quantitative relations between abnormal partitions of the lymph in semicircular canals (SSC) and idiopathic scoliosis (IS)?

Objective

Comparison of membranous SSC in IS and normal subjects.

Methods: We compared a scoliosis group SG (n=95, 75W. 20M.from 10 to 30 Y.) to a control group CG (n=33, 27W. 16 M., from 8 to 51Y.) Then, we divided SG in different subgroups: versus Cobb angle: SG1: 12 to 20°, SG2: 20 to 47°, versus curve type: thoracolumbar (TL), thoracic (T) and lumbar (L).

Data were acquired with a T2 MRI (G.E. Excite 1.5T, mode Fiesta). We processed the data with Brainvisa modules (http://brainvisa.info/) and novel automated modules permitting a scale jump and a partition between peri- and endo- lymph. Specific anomalies were scored for statistical study.

Results:

In SG, we found 43.1% of SSC aplasia parted on the lateral canals (56. 8 %), on posterior (44.2%), and on anterior (28.4%).

In SG, we also found abnormal communications between lateral and posterior canals in 54.7% and 3 patients without SSC anomaly.

In the CG, we found 9% of SSC anomalies located on the lateral canals (10.5 %), on posterior (9%) and anterior (7.5%) Mann-Whitney U test: sum of anomalies, posterior canals and anterior canals: $p<0, 0001$; lateral canals: $p=0.002$).

We compared SCC anomalies in the different SG subgroups:

-versus Cobb angle: no differences between the different groups (Mann- Whitney U test p= 0, 08),

- Versus curve type: no difference between the different groups ( Kruskal-Wallis test p= 0,69).

Discussion:

We found high percent of lateral canal aplasia & abnormal communications also affecting the lateral canal: these canals are known to be implicated in axial body rotation. These results could explain the results of Kramers in 2004 who observed an asymmetrical trunk rotation around the vertical axis and lateral deviations in IS patients of his study.

In the text, we bring the proof that abnormal connections also affect the bony canals. This result suggests a congenital origin: bony labyrinths are ossified early in utero. The additional fact that a large percentage of IS patients showed this specific anomaly, strongly suggests a possible genetic origin.

## Conclusion

Semi-circular canals anomalies detected in IS patients and confirmed by statistical analysis, may be a strong factor in the generation of idiopathic scoliosis. They could be detected and used in young children in case of familial scoliosis as a predictive factor of potential scoliosis.

## Introduction

The vestibular system is known to be a fundamental reference frame for the organisation of posture through vestibule-spinal reflexes and also through its role in the perception of spatial orientation. In former studies [1, 2], we have shown that subjects, with posterior basicranium asymmetry (PBA), have a spatial asymmetry of the vestibular organs and of the orbital cones, which are also often associated with central cerebellar asymmetry [3]. These patients exhibit a number of symptoms like ocular and head tilt, and are often afflicted with idiopathic scoliosis (IS)(See the companion paper).

Anomalies of the vestibular system have been suggested to be associated with scoliosis. In addition, Previc 1991[4], Bacsi 2004 [5] and Burwell 2006 [6] have also stressed the contribution made by vestibular asymmetry in postural disorders potentially leading to scoliosis. We shall present here new evidence suggesting that there may be a close causal relation between anomalies of the vestibular organs and scoliosis.

The peripheral vestibular system, namely the three semi circular canals(SSC) and the two otolithic organs (utricle & saccule) are embedded in the bony labyrinth, an enveloping structure, which contains and protects the membranous labyrinth, this latter being the sensitive part of the organ for equilibrium and space perception. This paper will only concern the study of semi circular canals anomalies: SSC are composed of ducts organised in three perpendicular planes interconnected with each other. They are filled with anatomically separated fluids: the peri- and the endolymph [7]. At the end of each canal is located an ampulla which contains ciliated cells responding to the displacement of the endolymph initiated by head movements: the bulk displacement of the lymph fluids causes the cilia of the cells to deform. Therefore, cells inputs are modulated by the shape of the canals and the knowledge of their internal shape is essential. Experimental studies on canals inputs revealed that each canal projects to a specific part of vestibular nuclei then to specific vestibulospinal tracts [8, 9, 10] with possible consequences on the rachis [5, 6] and vestibulo-ocular circuitry [11, 12, 13, 14]. Kushiro, recently, demonstrated that posterior canal (PC)-related vestibulospinal neurons projected their axons to C3 and L3 segment level. It is suggested that the PC-related vestibulospinal neurons form reflex arcs with antigravity muscles [15].

The aim of this study was therefore to compare the anatomy of the semi-circular canals in IS patients versus non scoliotic subjects. We used a novel validated and patented method based on MRI [16, 17, 18, 19] which serves as tool for characterizing anatomical anomalies of canals in vestibular patients [20].

## Material and Methods

The subjects of control and scoliotic groups were, previously submitted to the study establishing the relation between posterior basicranium asymmetry and scoliosis (see companion paper).

*Control group (CG)*

Justifying an MRI for healthy subjects was difficult; thus a control subject goup was recruited from patients undergoing maxillo-facial consultation and suffering from traumatic lesions of the temporomandibular joints(TMJ). An MRI of the temporomandibular joint can also capture the vestibular system because these two structures are close each other. T2 MRI has been carried through in the context of the treatment to verify the presence of blood or synovial anomaly inside the joints. None of these control subjects presented posterior basicranium asymmetry and their joint lesions were not spine-related. They were free of any spine deformation, whether congenital or acquired which was confirmed by rheumatologists of our team. For these reasons, they were included as “normal” control subjects. All of these subjects were consenting to give their MRI and spine data, in an anonymous way for the study.

This control group (**CG)** included 33 persons, 27 females and 16 males, ranging in age from 8 to 51 years with a mean age of 27 (std. dev: 5.6).

*Scoliotic group*

For the scoliotic group (**SC**), the exclusion criteria were to be free of acquired or congenital spine lesion, of visual or auditive malformation which could have postural incidences and of systemic illness. IS patients have been recruited from patients treated by physiotherapy and/or brace by rheumatologists of our team (hospital or clinics). The recruitment was time-randomized (23months): patients, presenting the inclusion criteria, were asked to participate to the study as they came along to consultations. MRI was prescribed, in the course of their therapeutic treatment to control anomalies in the skull base or cervical spine. They gave their consent to give their MRI and spine data, in anonymous way, for the study.
**SG** included 95 patients, 75 females and 20 males, ranging in age from 10 to 30 with a mean age of 17.3 (Std. dev.: 4.7). The greater number of females reflects the well-known large incidence of idiopathic scoliosis on female.

Their scoliosis was classified following two criteria:
1 –We measured the Cobb angle [21] according to the classical definition: "angle formed by the superior plateau of the superior limit vertebra and the inferior plateau of the inferior limit vertebra". The same operator measured manually the angles. Two sub-groups were distinguished according to the severity of the curve: SG1: from 12 to 20 degrees, including 57 subjects with a mean age of 14, 8 (std. dev.: 3.01), SG2: from 20 to 47° including 38 subjects with a mean age of 18, 9 (std.dev. 3, 85).
2- Location of the deformation along the spine according to the Cottalorda and Kohler classification (1997) [22]
The time-randomized recruitment, spread out over 23 months, allowed us to divide the same 95 patients in three sub-groups:

- 57 patients with a thoracolumbar deformation formed the TL subgroup (52 with left TL and 5 with right TL)
- 24 patients with a thoracic deformation formed the T subgroup (22 with right T and 2 with left T)
- 14 patients with a lumbar deformation formed the L subgroup (13 with left L and 1with right L)

*Modelling of the semi-circular canals*
We designed a completely novel method for visualizing the membranous labyrinth providing a "modelling" of the inner part of the vestibular canal ducts. This modelling is a 3D image reconstructed by the software and corresponding to the precise shape of the lymph moulded on the inside part of the membranous canal.

Data acquisition was performed with the EXCITE MRI (1.5T) system from General Electric using a head coil. A T2-weighted sequence in 3D FIESTA mode acquisition was used with the following variables: the orientation was axial, field of view, FOV=250.250mm, matrix dimensions MD=256.256, repetition delay TR=5ms, flip angle of 65 degrees, number of echoes NE=3, slice width SW= 31,25kHz, slice thickness SIT=1mm and the number of slices NSI= 192.
Steps of the program:
We processed the data using various modules of Brainvisa which is a free access software devoted to MRI available on the web(http://brainvisa.info/) and other novel automated modules.
The different steps of the process were the following:

1. Import and convert MRI data (Dicom format) given by the GE machine to 'Gis Brain visa' format. Data were stored in a file with the name of the patient.
2. Right and left cubic sub-volumes, containing the semi-circular canals images, were delimited in the initial volume in accordance with a procedure explained in the programme.
3. The modelling is based on the difference of luminosity between lymphatic fluids voxels and darker neighbouring structures voxels. The threshold of the endolymph luminosity (Brainvisa grey scale) is automatically indicated by the programme. With this process, we can isolate and extract the endolymphatic volume from the surrounding structures.
4. Anatomist 1.3, a Brainvisa module, is used to visualize the SSC models from every angle. The scale jump (see **figure 1A & 1B**) obtained with our process allows to visualize details of the 3D shape of the endolymph which do not normally appear clearly with T2 MRI reconstructions.

This new method allows the visualisation of the following anomalies of the semi-circular canal part of the membranous labyrinth;

(1) Filling defects of the membranous labyrinth, previously, described by Held as intra canal fibrosis [23]. This intra-luminal fibrosis can be a consequence of inflammation caused by various factors: infection, surgery or severe traumatism [24] (see **Figure 2**).

(2) Aplasia characterized by buds of canals, previously evocated by Adams in Charge syndrome [25] (see **figure 3).** We also found canals dysplasia characterized by a partial filling of the canal. Some dysplasias correspond to an intra-luminal obstruction (i.e.fibrosis), the others to a narrowing of the bony canal linings. The difference between these two options can be made by comparison of the MRI modelling to a CT scans modelling because CT reveals the bony shape of the canals.
Both of these techniques provide data in Dicom format which can be treated with our programme.

(3) Abnormal intercanal connections always found between lateral and posterior canals. These anomalies are here described for the first time (see **Figure 4).**

In order to obtain a general assessment of this pathology, the different semi-circular canals anomalies (1), (2), (3) were scored for statistical analysis: one point was assigned for each of the six semi-circular canals(three on each side of the head) giving a potential maximum of 6 points if all canals showed anomalies. A maximal score for SG patients was therefore 570 (the sum of 6 anomalies for 95 patients) and a score of 198 for CG patients (the sum of 6 anomalies for 33 subjects). Percentages of abnormal connections were also calculated for both groups.

Because inputs coming from lateral, anterior or posterior canals interfere differently on the muscular system through the vestibule-spinal system [8, 9, 10, 15], we also evaluated the anomalies of the three canals as follows: we combined the presence of SSC abnormalities between right and left side for each of the three canals. This method of quantification gave, respectively, for each pair of canals, a maximum of 190_(2x95) in SG and 66(2 x 33) in CG.
We, then, compared the scores for SC versus CG (in percentage) and also compared the scores in the SG sub-groups: i) SG1 vs. SG2, ii) T.vs. TL..vs. L. curve types. These results were also expressed in percentages.

## Statistical analysis

All statistical analysis was performed using SAS software (SAS Institute Inc. Cary, NC 25513) [26]. Data were expressed as mean, standard deviation and range.
P values<0.05 were considered statistically significant. For numerical variables (sums of counted canals anomalies), comparisons of groups involved various statistical analysis. The assumption of equal variances was tested using Fisher test: we used unpaired student t-test if the assumption was not rejected.

In other cases, we employed the test using unequal variances and Satterwaite's approximation. For categorical variables (SG versus CG and SG sub-groups) the comparisons were performed using the chi-square test or the Fisher's exact test if required. Scoring of SSC anomalies, in the case of two groups were compared using the Mann- Whitney U test and in case of more than two groups, we needed to use the Kruskal- Wallis.

## Results

*Comparison between SG and CG scored SSC anomalies*
In the **CG** group, 9 %( 18/198) subjects presented anomalies which were stenosis, intra-canal fibrosis and dysplasias. They were found in 10, 5% (7/66) in lateral canals, in 9 %( 6/66) posterior canals anomalies and in 7, 5% (5/66) in anterior canals.
In **SG** group, we detected a sum of 246/570 anomalies (43.1%) which were grouped as follows:

1) Anomalies were found in 56, 8% (108/190) in lateral canals, and 44, 2% (84/190) in posterior canals. The lowest percentage was for the anterior canals anomalies with 28.4 % (54/190).
We noted an important difference between SG and CG: in SG, the anomalies were severe. We discovered numerous cases of aplasia chiefly located on the lateral canals: 40% (76/190). The other lateral canal anomalies were large dysplasias 10,5% (20/190). And, we also found 0,01% (2/190) cases of aplasia on posterior canal and none on the anterior canal. *In CG, aplasias were totally absent*. We will discuss these findings.

2) A remarkable other finding was the presence of 54.7 %( 52/95) of abnormal connections in the scoliotic group (**Figure 5A, B, C**). *In CG, abnormal connections were totally absent.* It should be noted that in SG group, we found 3 patients without any canal anomaly. These patients, respectively, presented a12° lumbar curve, a 20° thoracolumbar curve and a 14° thoracolumbar curve.

The Mann-Whitney U test established the validity of the differences between SG and CG as following:
Concerning the combined scored anomalies of the semi-circular canals in SG versus CG: p<0, 0001
Concerning the scored anomalies in posterior canals or anterior canals in SG vs. CG: p<0, 0001
Concerning the scored anomalies in lateral canals in SG vs. CG: p=0.002.

*Comparison of scored SSC anomalies in SG subgroups*

We, then, proceeded to relate the anomalies with the severity of the curvature of the spine as measured by the Cobb angle. This analysis was made following the separation of the scoliotic group in subgroups as mentioned in the method. A first analysis concerned the relation between the number of anomalies and the Cobb angle for the scoliosis.

In SG1 (57 patients with Cobb angle between 12 and 20°), the study revealed 138 cases of SSC anomalies out of 342 (6 SSC anomalies for 57 patients) giving a percentage of 40, 3%.

In SG2 (38 patients with a Cobb angle greater than 20°), we found 108 SSC anomalies for a sum of 228(6SSC for 38 patients) giving a percentage of 47, 3%

Statistically, using the Mann- Whitney U test, these results were not different (p= 0, 08)

A second analysis concerned the relation with the type of spinal curvature

In TL (57 patients with thoracolumbar deformation), the study revealed 146 cases of SSC anomalies for a sum of 342(6x57) giving a percentage of 42, 6%

In T (24 patients with thoracic deformation), the scored results were of 64 cases for 144, giving a percentage of 44, 4%

In L (14 patients with lumbar deformation), we found 36 cases for a sum of 84(6x14) giving a percentage of 42, 8 %.

The Kruskal-Wallis test revealed no significant difference (p=0, 69) between these different categorical subgroups.

## Discussion

Our new method of visualization of the membranous labyrinth revealed a number of anomalies both in control and scoliotic subjects

Anomalies identified in the control group (CG) were stenosis and /or intra-canal fibrosis which "could be" a consequence of traumatism or infection. As suggested by different authors [23, 24, 25], the membranous labyrinth is very fragile and responds to aggressions by intra lumen fibrosis. In the methods, we noticed that CG patients suffered of acquired temporo-mandibular lesions (TMJ). CG stenosis and intra-canal fibrosis, were probably acquired at the time of the TMJ damage caused by infection or traumatism.

The main result of this study concerning scoliotic patients is that SSC anomalies found in SG (43, 1%) were highly superior in number to CG (9%). They were also severe: aplasia of canals, abnormal inter-canal connections. As mentioned in methods section, SG patients were submitted to a previous study (see companion paper) which confirmed the strong relationship between their posterior basicranium asymmetry and scoliosis

Thus, we can suggest that SCC anomalies could be congenital. Our suggestions are consistent with those of Kwaks (2002)[27] and Philips (2004)[28] who demonstrated, in a experimental study on zebra fish, links between cerebellum abnormalities and structural anomalies of inner ears [4, 5]. More, the report by Burwell (2006) [8] has suggested that neurodevelopmental anomalies, in Humans, are implicated in the CNS body schema. The consequences of these anomalies are disturbances of the coordination of movement and posture by the vestibule-spinal and cerebellar outputs. Child growth could reveal the effects of these neurodevelopmental anomalies on spine growth as evocated by Yamada [29] in 1984;

In SG, we found a high percentage of <u>lateral </u>canal anomalies (43%) and abnormal connections between the <u>lateral</u> and posterior canal (54, 7%). These results could be pointed out because lateral canals are physiologically controlling head and body rotation in yaw. This might explain the results of Kramers in 2004[30] who examined scoliotic subjects during walking in order to identify asymmetries which may be related to a neurological dysfunction and found a normal physiological pattern of hip, knee and ankle in sagittal plane but an asymmetrical trunk rotation around the vertical axis and lateral deviations.

We also found abnormal connections, never described before this study. We compared two modellings of a same case of connection: one obtained by MRI and the other obtained with CT scans. This latter highlighted the same abnormal connection on bony external canals. This result is very important because it gives the proof of a congenital origin of this anomaly (bony labyrinths are ossified early in utero). The additional fact that a large percentage of IS patients showed this specific anomaly (54, 7%), strongly suggests a possible genetic origin of abnormal connections.

However, three patients reported no canal anomaly confirming the multifactorial origin of IS. One of these other possible origins could be linked to the otolithic system which is chiefly devoted to linear acceleration, including gravity and highly implicated in postural adjustments as heading [31]. Unfortunately, anomalies of the otolithic system cannot, presently, be explored by imaging and physiological tests are still treated with caution. However, we have mentioned that SG patients were affected with posterior basicranium(PBA) asymmetry. This last anomaly involves spatial asymmetry of the otolithic organs because they are embedded in PBA and allows the hypothesis of a participation of asymmetrical otolithic inputs in IS [1, 2]

*Canals anomalies versus severity of the curve*:
The absence of differences between SG1 & SG2 may be taken as evidence against our suggestion that semi-circular canls and asymmetries are contributing to idiopathic scoliosis. However, the mean age of the patients in these two groups was, respectively, 14.8 and 18.9. In SG1, patients were younger and IS were still evolutive. So canal anomalies could be considered as an early factor. Further linear study of scoliosis evolution versus canals anomalies should shed some light on this question. In addition, the complexity of the regulation of the posture of the spine excludes any simple relationship between canals anomalies and Cobb angle values.

*Canal anomalies versus scoliosis types*:
No significant differences were detected between the various types of spinal deformities (thoracolumbar, thoracic or lumbar curve) and the percentage of SSC anomalies. Recently, Kushiro[15] has highlighted that SSC inputs project to specific zone of the spine (for example, posterior SSC project to C3 and L3). Thus we should, now, study thoroughly the location of SSC anomalies versus the location of spine deformation.

Finally, we have in previous papers suggested that influence of vestibular asymmetries or dysfunction is not necessarily only by direct vestibule-spinal mechanisms but can also be mediated through the cerebellar control of posture and even also by higher descending cognitive processes involving the perception of the body schema [32] Therefore, although our findings clearly show the co-occurrence of canals anomalies and mild idiopathic scoliosis, the causal chain of events linking the two still has to be established. The potential importance of asymmetries of the orbital cones on the proprioception of the eyes which is known to have a powerful effect on postural control will also have to be considered [2, 3];

## Conclusion

Semi-circular canals anomalies detected in IS patients and confirmed by statistical analysis, may be a strong factor in the generation of idiopathic scoliosis. They could be detected and used in young children in case of familial scoliosis as a predictive factor of potential scoliosis.

Titles & Legends of Figures

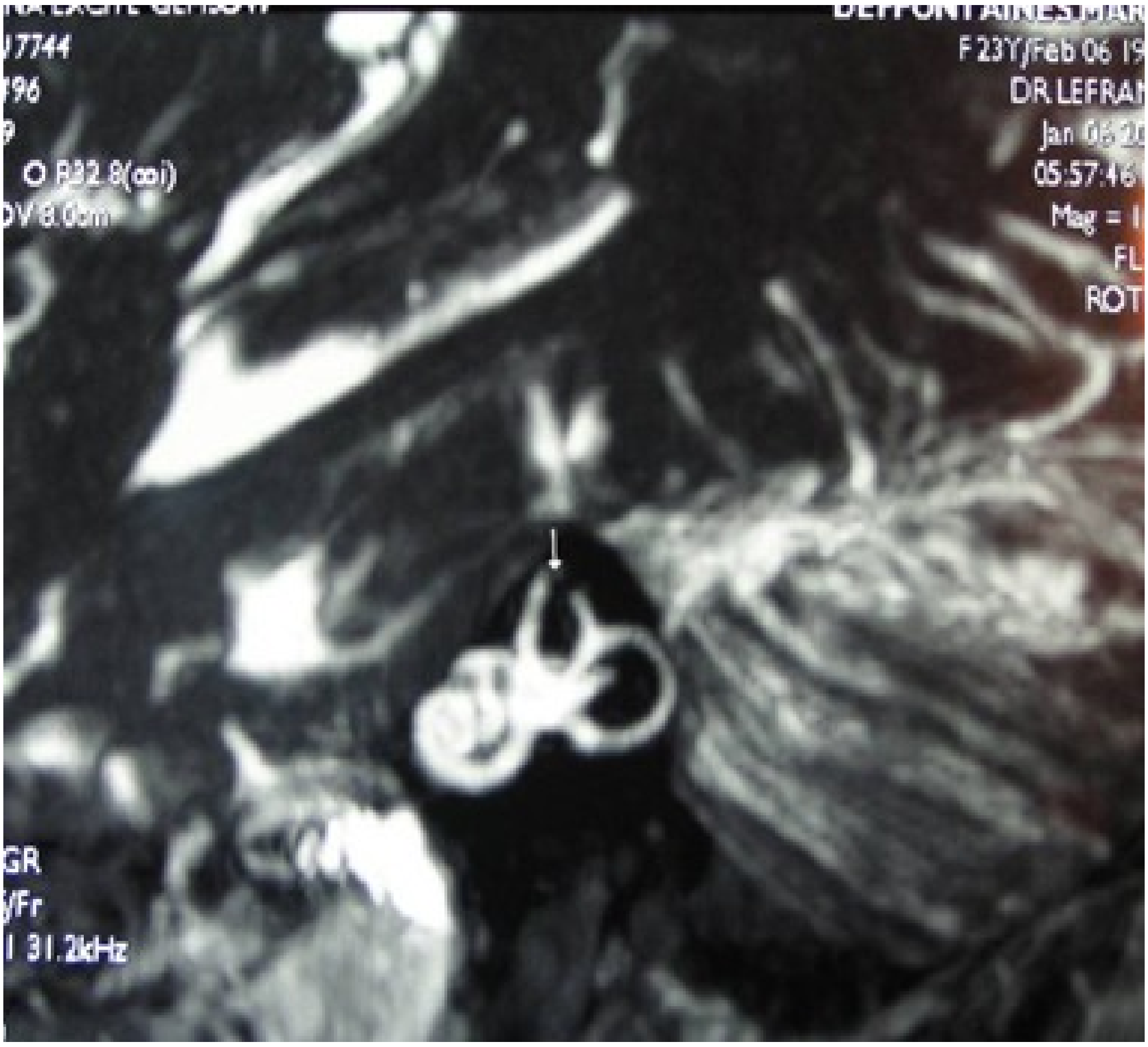

Figure 1A
**Visualization of the semi-circular canals obtained with MRI (G.E. Excite 1.5T.)**
MRI T2 allows a virtual extraction of lymphatic fluids out of the bony canals but the scale is too small for examining the details of the ducts. The white arrow shows a discontinuity of the fluid of the anterior canal which indicates an obstacle inside the duct.

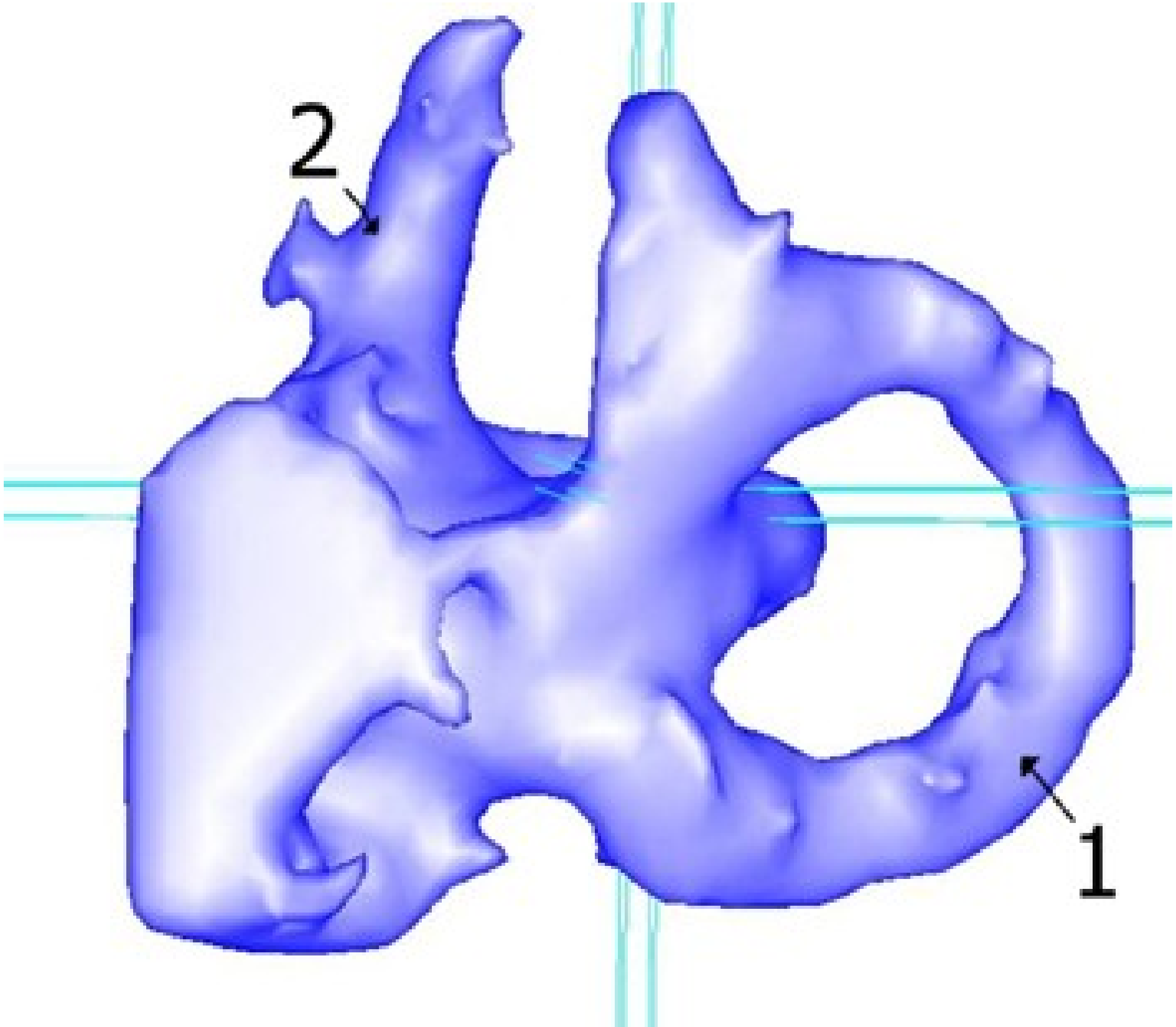

Figure 1B

**The same semi-circular canals visualization obtained with our novel modelling process**

One observes the scale jump obtained with the novel process. The orientation of the structure is identical to 1A. The discontinuity of the anterior canal fluid appears identical but more precisely (1: anterior canal, 2: posterior canal).

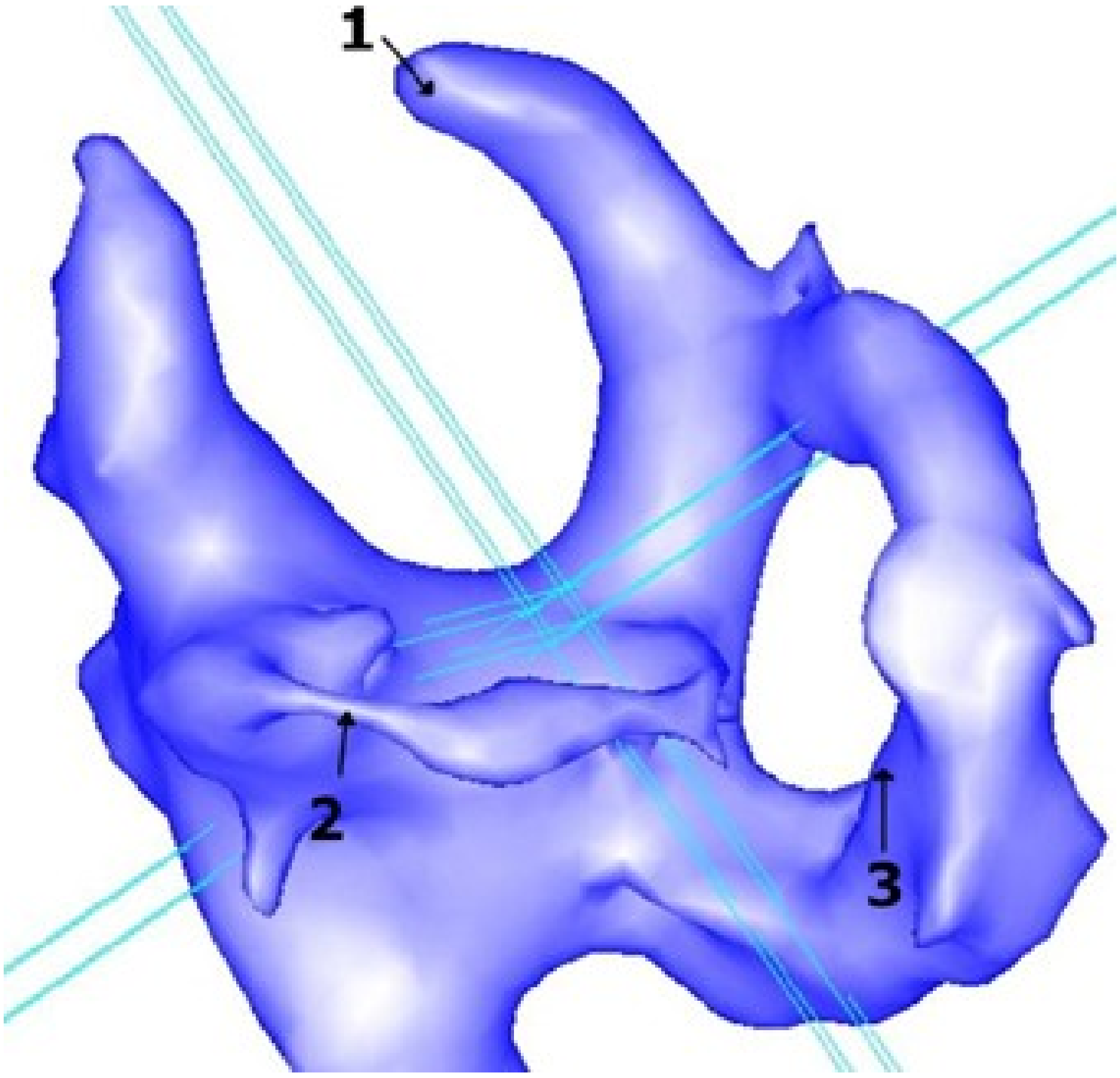

Figure 2

**Modeling of semi-circular canals after a tremendous head impact**

The traumatism induced intra-canal fibrosis; 1: anterior canal presenting a large fibrosis,
2 lateral canal partially blocked by the fibrosis, 3: non injured posterior canal.

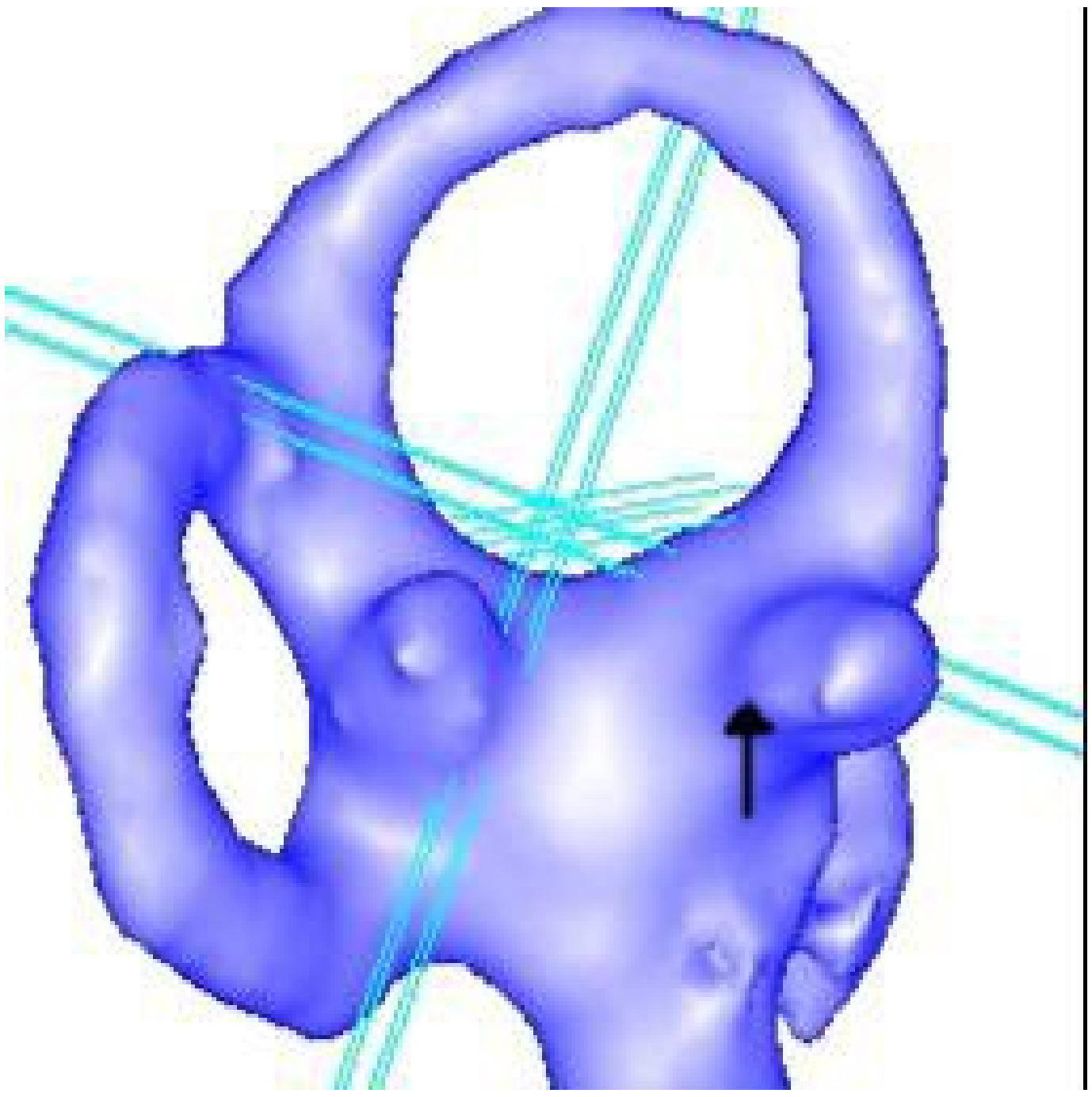

Figure 3

**Canal modelling of IS patient suffering from lateral canal aplasia**

The black arrow shows an important aplasia of the lateral canal. One observes canal buds. This anomaly is congenital and frequent in IS (56.8%of cases of our study)

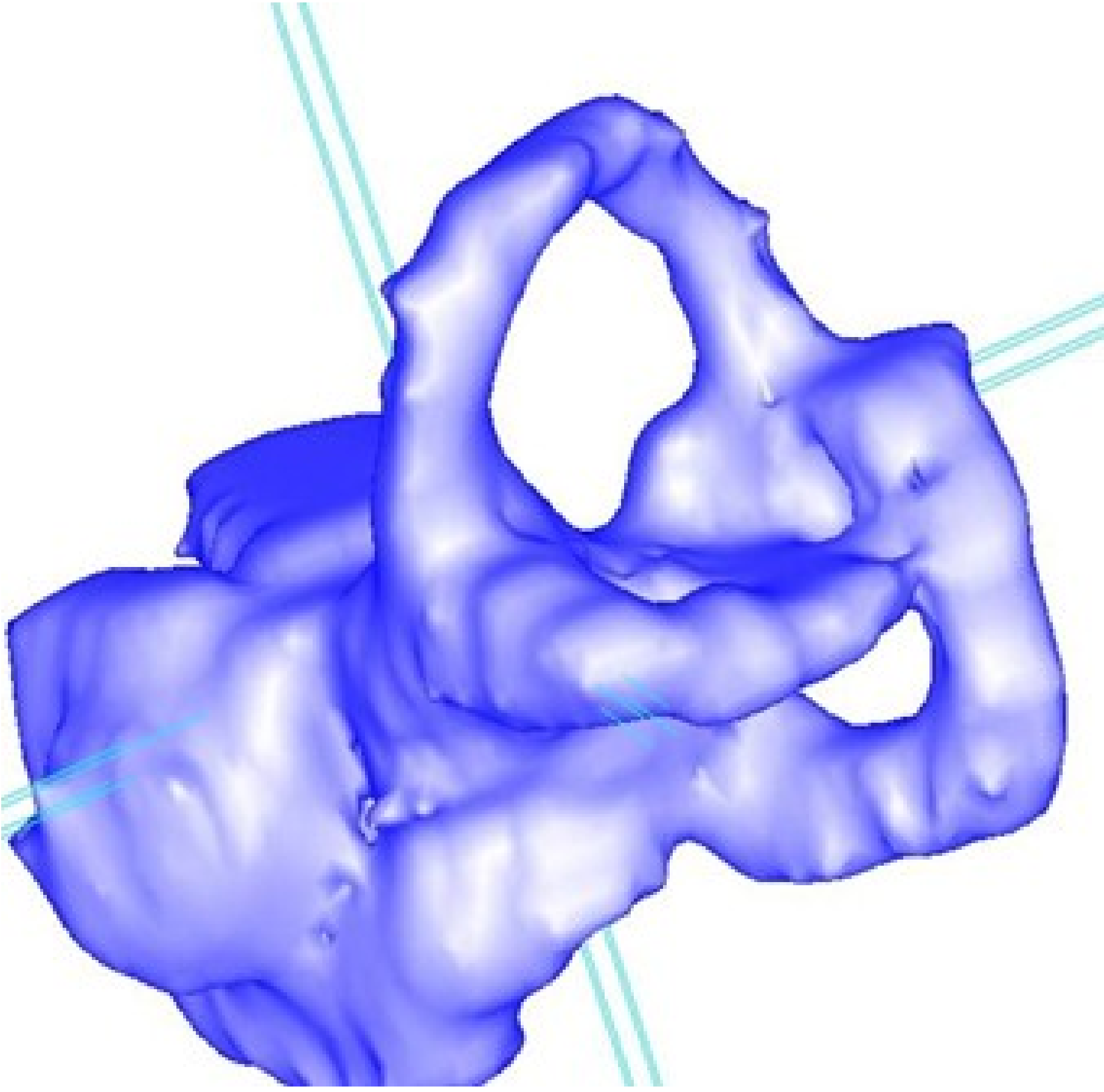

Figure 4

**Modeling of inter-canal communication**

This anomaly was never described. One observes an abnormal communication between the lateral and the posterior canal. The anomaly was always located between these two canals. In our study, this anomaly seems specific of IS: 54.7% of cases of our study and no control subject (n= 33) affected with.

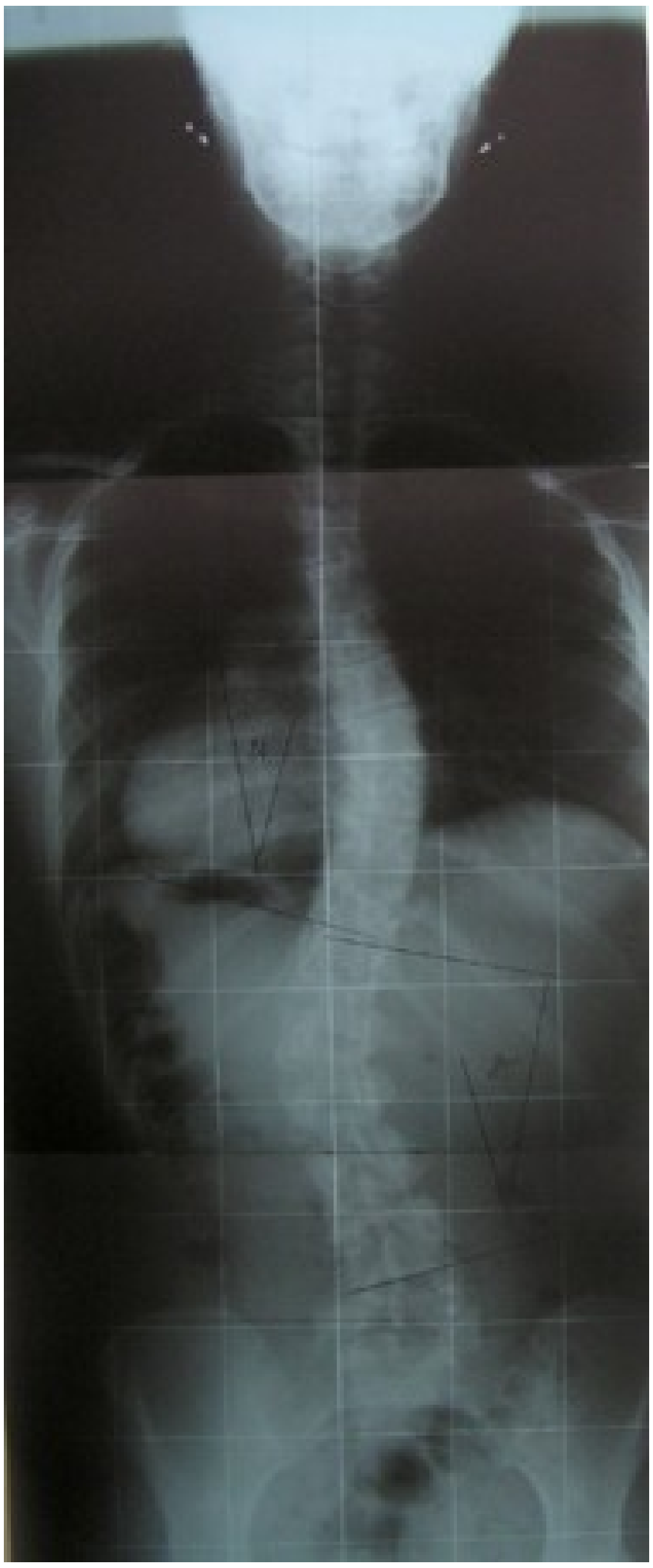

Figure 5A
**Scoliosis X-Rays**
This 14 Y. old girl was affected with an important IS thoraco-lumbar scoliosis (thoracic angle =24°, lumbar angle=25°).

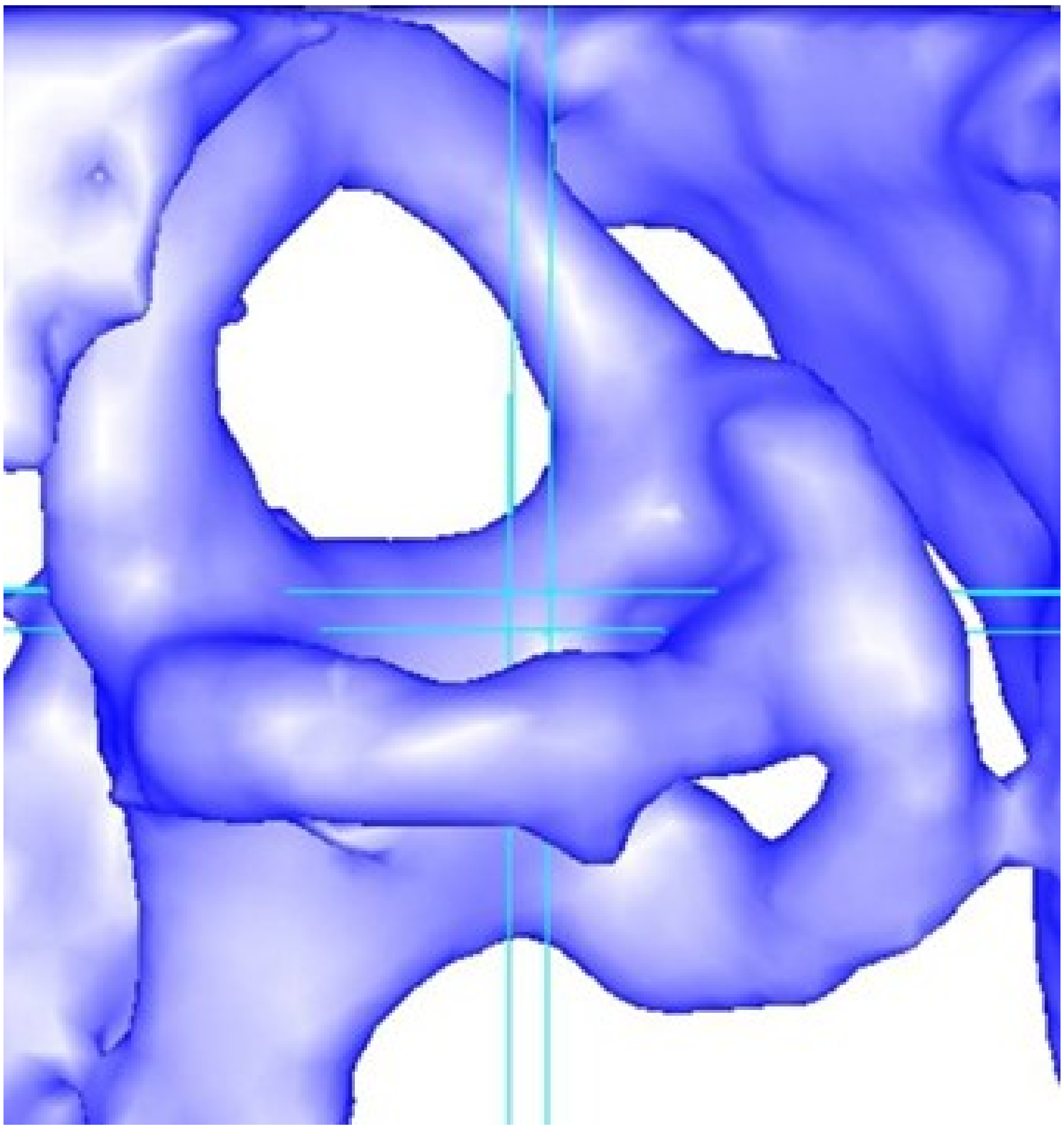

Figure 5B

**Modeling of her right semi-circular canals**

The modeling of the right canals shows a large abnormal communication between the lateral and posterior canal.

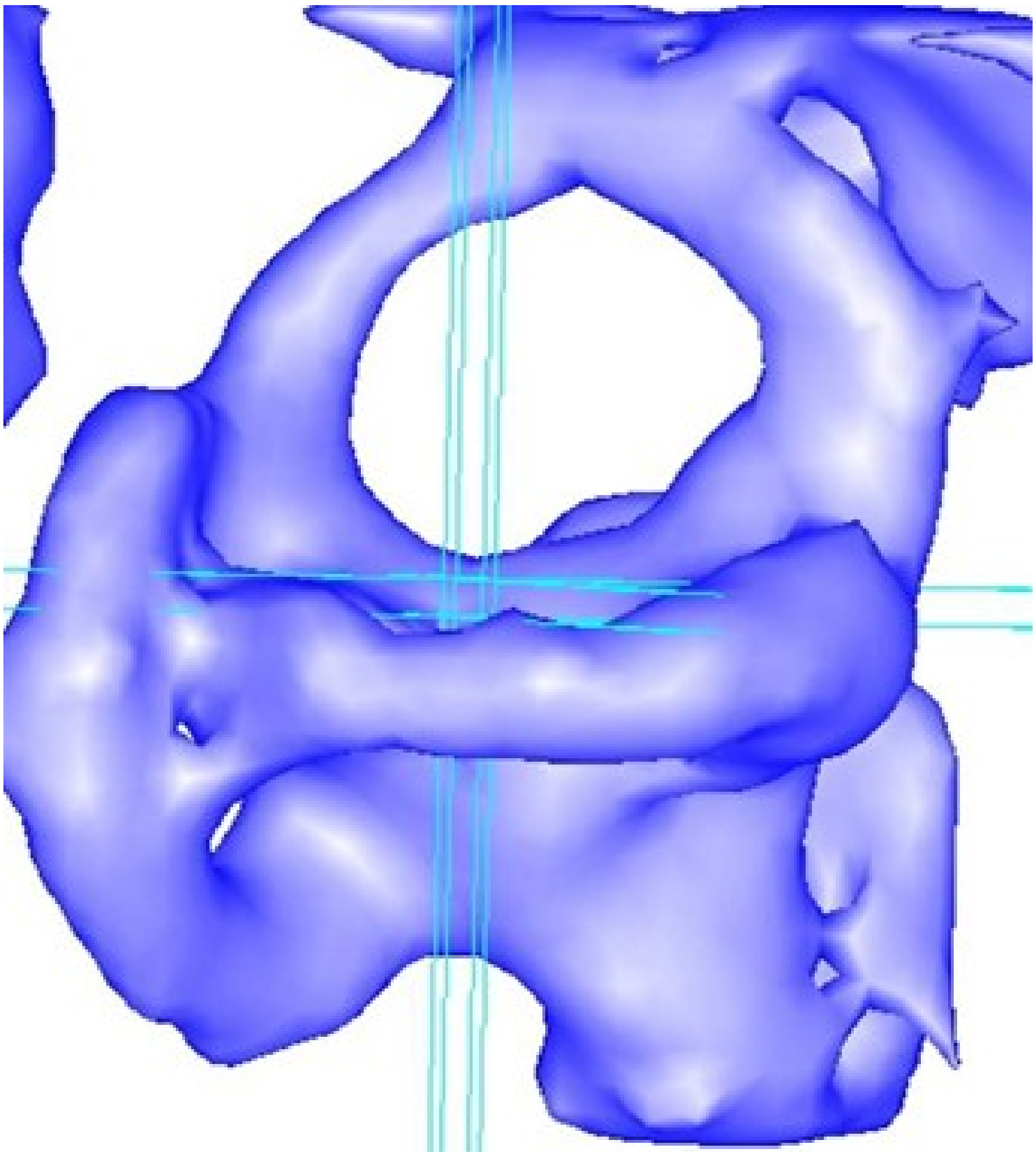

Figure 5C

**Modeling of her left semi-circular canals**

The modelling shows the same anomaly on the left side. In our study, in case of bilateral communications, all of our patients were afflicted with a severe double convexity deformation. The inverse was not observed**:** we found some IS thoraco-lumbar scoliosis without this abnormal double communication.